**Dual-switch control of a layer-locked anomalous valley Hall effect in a sliding ferroelectric antiferromagnet**

Quan Shen[1], Wenhu Liao[1], Degao Xu[2], Jiansheng Dong*[1], and Jianing Tan*[1]

[1]*Department of Physics, Jishou University, Jishou 416000, Hunan, China*

[2]*School of Physics and Optoelectronic Engineering, Yangtze University, Jingzhou 434023, China*

*Corresponding authors: jsdong@jsu.edu.cn, jianing_tan@163.com

**ABSTRACT**

The integration of ferroelectric (FE) and antiferromagnetic (AFM) orders in two-dimensional (2D) materials provides a promising avenue for the nonvolatile control of coupled spin and valley degrees of freedom, a capability central to advancing spin-valleytronics. However, realizing a single material system where these quantum states can be independently and reversibly manipulated by distinct stimuli, a prerequisite for multifunctional devices, has remained elusive. Here, we demonstrate a dual-switch mechanism in bilayer $VS_2$, a room-temperature FE-AFM system, that enables electrical and magnetic control of a layer-locked anomalous valley Hall effect (AVHE). First-principles calculations reveal that interlayer sliding breaks spatial inversion symmetry, inducing a switchable out-of-plane FE polarization that coexists with interlayer AFM. The spin-orbit coupled valley polarization can be reversibly switched either by FE polarization reversal or by a magnetic-field-induced spin-flip transition, confirming the existence of electrically and magnetically addressable valley states. The Berry curvature exhibits both valley-contrasting and layer-locked characteristics, which underpin a switchable Hall response. Notably, electric and magnetic switching are functionally equivalent in modulating valley, layer, and spin indices, revealing strong magnetoelectric coupling. This work establishes a multidegree-of-freedom operational paradigm in 2D multiferroics and opens a viable design pathway toward multi-state memory and spin-valleytronic logic devices.

## I. INTRODUCTION

The active control of spin and valley degrees of freedom, complementary information carriers in condensed matter, lies at the heart of spintronics and valleytronics [1,2]. While spin manipulation underpins phenomena from magnetoresistance to spin-transfer torques [3-5], valleytronics seeks to exploit the character of carriers at distinct momentum-space extrema [6,7]. A fundamental objective in this pursuit is to lift valley degeneracy to generate measurable, valley-polarized [8,9]. Coupling these valley states with magnetic order presents a promising pathway to engineer correlated spin-valley states, offering a rich platform for exploring emergent quantum phenomena and novel device concepts.

The discovery and characterization of compensated magnetic materials, specifically, antiferromagnets (AFMs) and, more recently, altermagnets (AMs), have redefined the landscape of spintronics [10-15]. While traditional ferromagnets (FMs) exhibit net magnetization, their susceptibility to external fields and slow spin dynamics limit device scalability and speed. In contrast, compensated magnets offer robustness against magnetic perturbations and support terahertz-frequency spin dynamics due to their vanishing net moment. As a distinct class of magnetic materials, AMs bridge a key gap by exhibiting momentum dependent spin splitting akin to FMs while retaining zero net magnetization [16,17]. However, a fundamental limitation persists: AFMs can host spontaneous valley polarization but lack intrinsic spin polarization, whereas AMs possess a spin-split band structure but require external stimuli for valley polarization [18]. Therefore, the development of a single AFM system that intrinsically unites spin and valley degrees of freedom remains a pivotal, and yet unrealized, objective for next-generation spin-valleytronic technologies.

Ferroelectric antiferromagnets (FE-AFMs) provide a unique platform for integrating spin and valley polarization within a single material. These systems combine spin-polarized bands, zero net magnetization and spin splitting, thereby hosting both spin and valley polarization within an AFM lattice [19,20]. Crucially, they retain the inherent advantages of AFMs, such as immunity to external magnetic fields, negligible stray fields, and terahertz spin dynamics, while also enabling non-volatile electric-field control via FE switching. This integration offers a versatile foundation for multiferroic manipulation of multiple degrees of freedom, opening pathways to coupled electronic phenomena in two-dimensional (2D) materials and informing the design of next-generation spintronic and valleytronic devices.

In such systems, crystal symmetry breaking gives rise to a Berry curvature that underpins various Hall effects [21-24]. Specifically, the breaking of time-reversal ($\mathcal{T}$) symmetry leads to the quantum anomalous Hall effect [25-28], which is tied to spin polarization, whereas the breaking of spatial inversion ($\mathcal{P}$) symmetry underlies the valley Hall effect [29-31]. In layered materials, the stacking geometry introduces a layer degree of freedom, which enables the Berry curvature to be encoded in the combined spin, valley, and layer indices. This gives rise to a set of distinct transport phenomena, including spin, anomalous, valley and layer Hall effects, with the latter two representing especially promising frontiers [32-34]. A central challenge, however, lies in 2D AFMs: the combined $\mathcal{PT}$ symmetry often forbids the simultaneous emergence of valley and layer Hall responses [35-37]. Conventional external fields are limited to manipulating individual degrees of freedom sequentially, which is insufficient for the correlated and simultaneous control demanded by multifunctional quantum devices. Consequently, the achievement of synergistic valley and layer Hall effects in compensated magnetic systems remains an outstanding challenge.

In this study, we use first-principles calculations to investigate a bilayer $VS_2$

system engineered as a sliding FE-AFM, which yields a coupled state of interlayer AFM order and switchable out-of-plane FE polarization. We demonstrate that, within this platform, the spin and valley polarization is reversible and bidirectional; it can be flipped either by sliding-mediated FE switching or by magnetic moment reversal. The computed Berry curvature exhibits valley-contrasting and layer-locked characteristics, establishing a fully switchable, layer-locked anomalous valley Hall effect (AVHE). Crucially, FE polarization reversal and magnetic switching are functionally equivalent in governing spin, valley, and layer transport, a direct manifestation of pronounced magnetoelectric coupling. This equivalence suggests that electrically switchable spin-valley phenomena may emerge universally in layered FE-AFMs with broken $\mathcal{PT}$ symmetry. Such systems thus provide a versatile conceptual and material platform for electric- and magnetic-field-controlled valleytronic and spintronic devices.

## II. COMPUTATIONAL METHODS

Density functional theory (DFT) calculations were performed using the Vienna *ab initio* Simulation Package (VASP) [38,39]. We employed the projector augmented-wave (PAW) method and the Perdew-Burke-Ernzerhof (PBE) generalized gradient approximation (GGA) to describe the exchange-correlation functional [40]. Structural relaxations were considered converged when the total energy change and maximum atomic force fell below thresholds of $10^{-6}$ eV and 0.001eV/Å, respectively. A plane-wave kinetic energy cutoff of 500 eV was used. A vacuum region of 20 Å was applied perpendicular to the surface to suppress spurious interactions between periodic images. To account for van der Waals interactions in the bilayer system, Grimme's DFT-D3 correction was applied [41]. The bilayer configuration was generated by sliding the top layer relative to a fixed bottom layer. All atomic positions were then fully relaxed, and the total energy of each configuration was calculated. For the monolayer and bilayer systems, Brillouin zone integration was performed using a Monkhorst-Pack *k*-point grid centered at the $\Gamma$ point with a density of $30\times30\times1$. To account for the strong electron correlation in the localized *d*-orbitals of V, we employed the DFT+U approach using the Dudarev formalism with an effective Hubbard parameter of $U_{eff}=3$ eV [42,43]. The dynamic properties and thermal stability of the structure were assessed using a $5\times5\times1$ supercell. Phonon spectra were computed using the PHONOPY code, complemented by separate *ab initio* molecular dynamics (AIMD) simulations [44,45]. Wannier functions for the S *p* and V *d*-orbitals were constructed using WANNIER90. Subsequent Berry curvature calculations were also performed with this package [46]. The minimum-energy pathway and associated slip potential barrier were determined using the climbing-image nudged elastic band (CI-NEB) method [47]. FE polarization was computed via the Berry-phase approach [48,49].

## III. RESULTS AND DISCUSSION

### A. Stability and magnetism in monolayer

The optimized monolayer $VS_2$ structure exhibits trigonal symmetry (space group *P*3m1, No. 156) with an in-plane lattice constant of 6.47 Å. The monolayer $VS_2$ adopts a hexagonal lattice with a vertical S-V-S stacking sequence [Fig. S1(a)]. Analysis of the electron localization function (ELF) reveals charge density localized primarily around the S anions, indicating a predominantly covalent character for the V-S bonds [Fig. S1(d)]. Phonon spectrum calculations and AIMD simulations confirm the monolayer dynamic and thermal stability [Figs. S1(b) and S1(c)). The total energy difference between FM and AFM configurations of monolayer $VS_2$ was calculated using the GGA+U method to determine the magnetic ground state [Figs. S2(a) and S2(b)). Consistent with established values, we used $U_{eff}=3$ eV for V [19], which yields an energy difference of $\Delta E$=468 meV, confirming the FM state as the magnetic ground state of monolayer $VS_2$. The electronic structure without spin-orbit coupling (SOC) reveals an indirect bandgap of 0.48 eV [Figs. S1(e) and S1(f)]. Both the valence band maximum (VBM) and conduction band minimum (CBM) near the Fermi level originate from the same spin channel, indicative of a unipolar magnetic semiconductor. The spin-resolved density of states is asymmetric [Fig. S1(f)], confirming a net spin polarization consistent with a FM ground state.

**B. Magnetoelectric coupling of ferroelectric antiferromagnet bilayer**

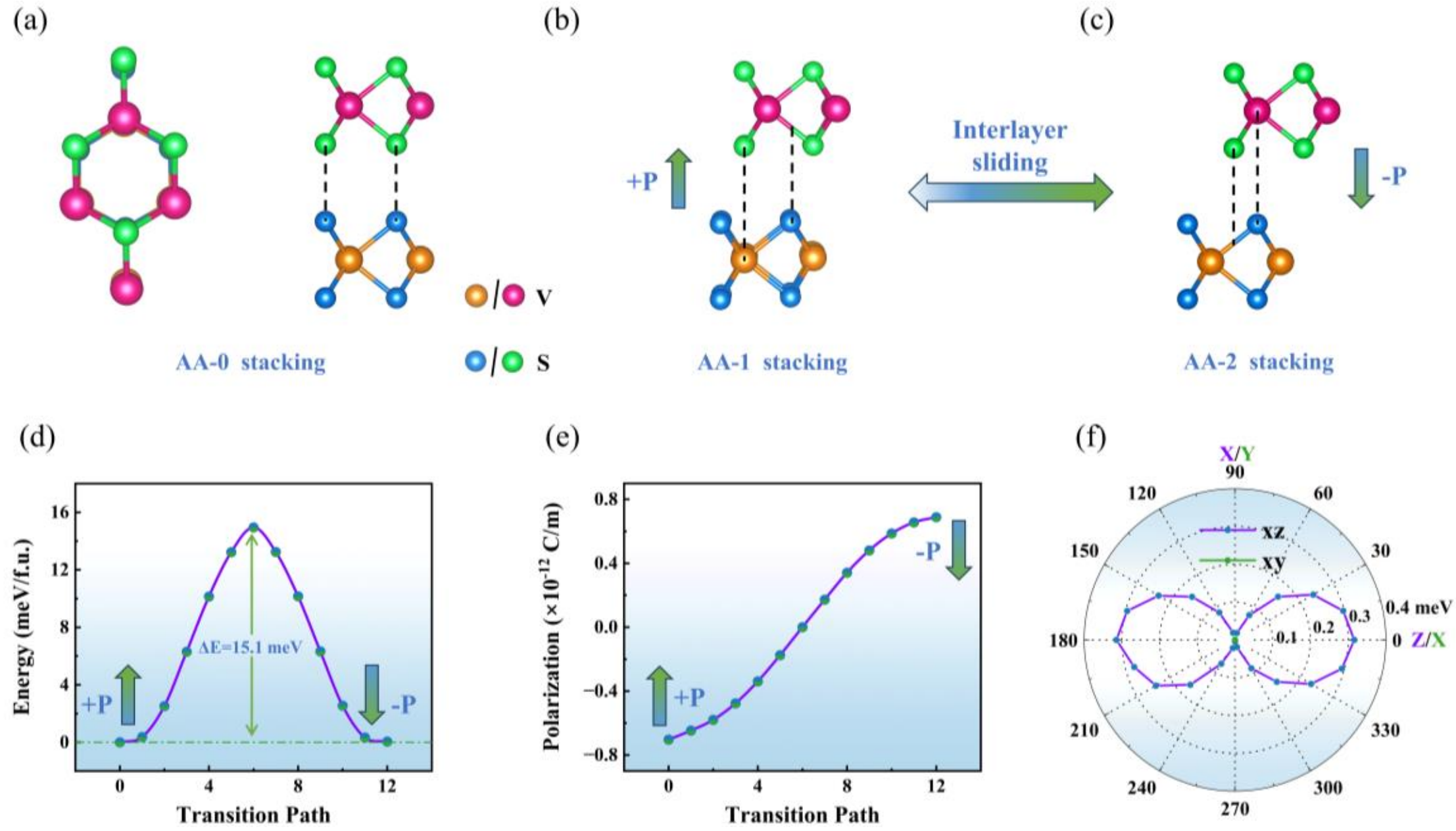

FIG. 1. Atomic schematics of three distinct AA-stacking configurations: (a) AA-0, (b) AA-1, (c) AA-2. V atoms in the upper and lower layers are represented by pink and orange spheres, respectively; S atoms are shown as green (upper) and blue (lower) spheres. (d) Sliding energy barrier for interlayer translation, corresponding to a switching pathway from the FE polarization-up (AA-1) to the FE polarization-down (AA-2) state. (e) Evolution of the out-of-plane FE polarization magnitude along the switching pathway. (f) MAE of the AA-1 configuration, showing the energy dependence on the magnetization orientation.

The stacking of 2D magnetic materials can induce emergent electronic and topological phenomena, such as exemplified by the layer-tunable topological phases in $MnBi_2Te_4$ [50], the type-III magnetoelectric coupling in bilayer $MnPSe_3$ and the transition of AFM monolayers into an AM ground state [51]. However，the spin splitting in bilayer AFMs is symmetry-locked and difficult to control via external electric or magnetic fields, creating a key obstacle to their application. In most monolayer crystals, in-plane lattice symmetry prohibits a net out-of-plane FE polarization. This constraint, however, can be circumvented through stacking: specific lateral displacements between layers break $\mathcal{P}$ symmetry, generating a switchable out-of-plane dipole via the so-called sliding FE mechanism. This form of switchable FE polarization has been experimentally demonstrated in a range of 2D systems, including h-BN [52,53], InSe [54,55], and $MoTe_2$ [56,57], providing a generic pathway to engineer electrical control in otherwise non-polar materials.

Here, we investigate how bilayer stacking configurations influence the properties of $VS_2$, analyzing the distinct electronic and magnetic behaviors that emerge from different interlayer arrangements. The bilayer $VS_2$ exhibits two principal stacking motifs: AA and AB, each with three distinct stacking variants (labelled -0, -1, -2; Fig. S3). We computed the total energies of different magnetic configurations for all six stacking variants. The results consistently show that the AFM state is energetically more favorable than the FM state [Fig. S4]. This establishes a robust preference for interlayer AFM coupling, thereby classifying bilayer $VS_2$ as an A-type AFM with intralayer FM and interlayer AFM order. Furthermore, the AFM configurations for three stacking variants, AA-1, AA-2, and AB-0, are energetically nearly degenerate, indicating the presence of multiple accessible metastable states [Fig. S4]. Within the AA series, the AA-0 stacking corresponds to a direct vertical alignment of monolayers, preserving mirror symmetry with respect to the $z$-axis [Figs. 1(a) and S3(a)]. This geometry, protected jointly by mirror and $\mathcal{T}$ symmetry, yields spin-degenerate bands in the absence of SOC [Fig. S5(a)]. In contrast, the AA-1 and AA-2 configurations are generated by translating one layer relative to the other along one of three equivalent in-plane directions, $t_1\left(\frac{2}{3},-\frac{1}{3},0\right)$, $t_2\left(-\frac{1}{3},-\frac{1}{3},0\right)$ or $t_3\left(\frac{1}{3},\frac{2}{3},0\right)$ [Figs. 1(b) and 1(c)].

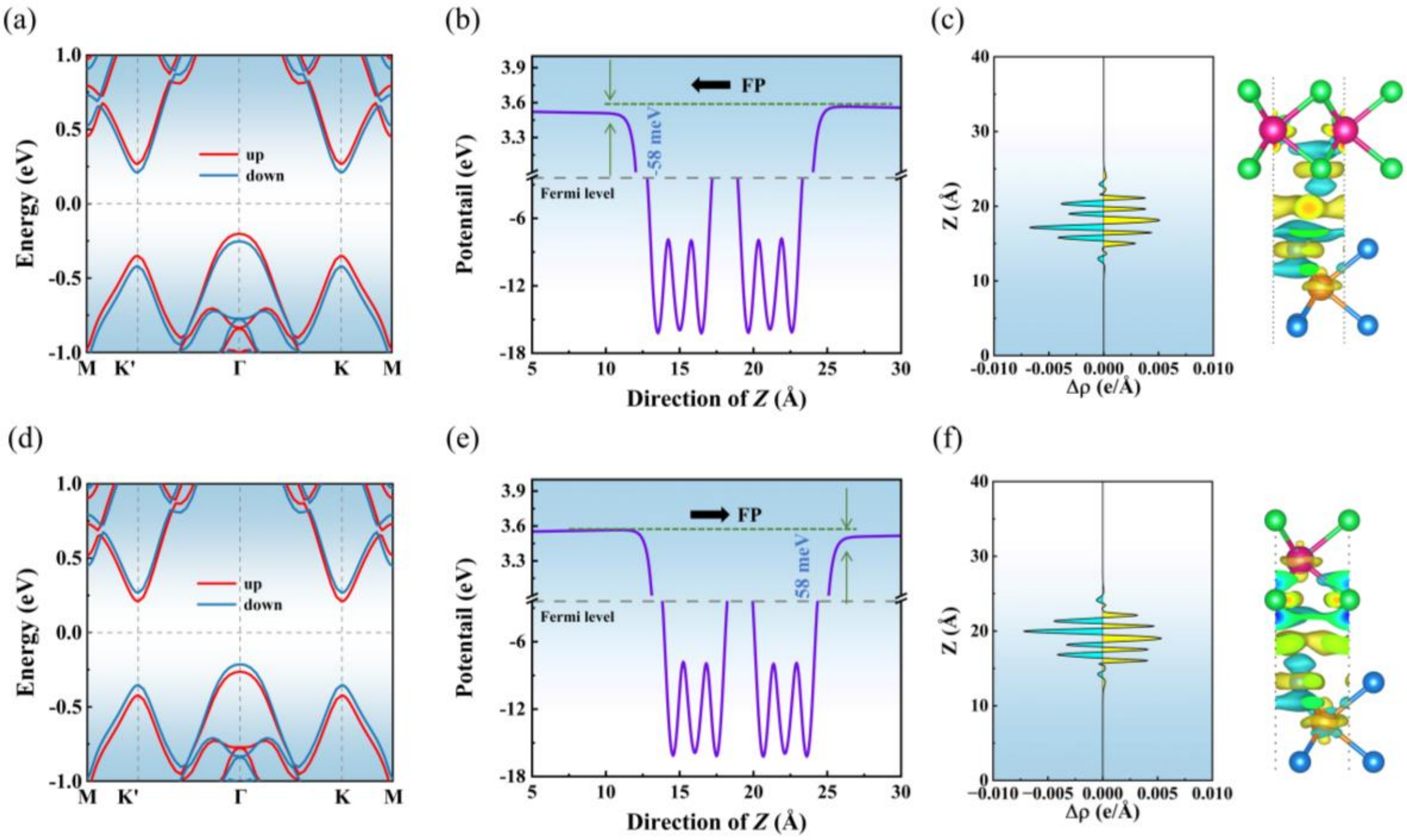

FIG. 2. (a,d) Calculated band structures for the (a) AA-1 (P↑) and (d) AA-2 (P↓) configurations. (b,e) Corresponding averaged electrostatic potential along the $z$-direction. (c,f) Screening charge distributions and differential charge density for the (c) AA-1 and (f) AA-2 states, illustrating the charge redistribution associated with FE switching. Yellow and cyan isosurfaces denote charge accumulation and depletion, respectively.

These sliding-stacked arrangements break the out-of-plane mirror symmetry, thereby lifting $\mathcal{P}$ and inducing a stable out-of-plane FE polarization. The resulting FE polarization vectors are perpendicular to the plane, oriented in the $+z$ and $-z$ directions for AA-1 and AA-2, respectively. The energy barrier for interlayer sliding between the AA-1 and AA-2 FE states, computed via the CI-NEB method, is only 15.1 meV/f.u. [Fig. 1(d)]. This low value, which smaller than those of many comparable bilayer sliding systems, confirms that the FE polarization is easily switchable. Calculations using the Berry-phase approach establish that the AA-1 and AA-2 states exhibit opposing out-of-plane FE polarizations of magnitude $0.69 \times 10^{-12}$ C/m. The FE polarization evolves linearly with interlayer sliding displacement, confirming its direct geometric origin [Fig. 1(e)]. Notably, this FE polarization exceeds previously reported values for known sliding FE, such as bilayer $YI_2$ ($0.26 \times 10^{-12}$ C/m) and $FeCl_2$ ($0.41 \times 10^{-12}$ C/m) [18,58]. For the AA-1 bilayer, magnetic anisotropy energy (MAE) calculations indicate an in-plane easy axis. The energy exhibits a pronounced dependence on the magnetization angle within the $xz$-plane, while remaining nearly invariant for rotations within the $xy$-plane [Fig. 1(f)].

All AB-stacked configurations possess $\mathcal{P}$ symmetry (centrosymmetric) and consequently lack a FE dipole. Their band structures are spin degenerate [Fig. S6], including the metastable AB-0 phase, which is therefore unsuitable for valley or layer polarized phenomena. Therefore, our analysis thus focuses on the non-centrosymmetric, energetically stable AA-1 and AA-2 configurations. The AA-1 and AA-2 configurations of bilayer $VS_2$ exhibit stable, out-of-plane FE polarizations of equal magnitude but opposite sign. This FE polarization stems from an asymmetric interlayer charge transfer, induced by a layer-dependent built-in electrostatic potential of $\pm 58$ meV (AA-1: $-58$ meV; AA-2: $+58$ meV) as evidenced by planar averaged potential, screening charge distributions, and differential charge density analyses [Figs. 2(b) and 2(f)]. In the absence of SOC, both states are indirect gap semiconductors with a bandgap of 0.42 eV [Figs. 2(a) and 2(d)]. A key finding is that interlayer sliding, which reverses the FE polarization (AA-1↔AA-2), concomitantly inverts the spin polarization. This robust magnetoelectric coupling directly links the FE order parameter to the spin texture, enabling integrated electrical control of entangled charge, spin, and valley degrees of freedom.

**C. Electrical and magnetic control of layer-locked AVHE**

From a symmetry perspective, the realization of a layer-locked AVHE hinges on three sequential conditions. $\mathcal{P}$ symmetry must be broken. When $\mathcal{P}$ symmetry is preserved, the spin-up ($K$) and spin-down ($K'$) valleys remain strictly degenerate in energy, and their Berry curvatures satisfy $\Omega(K)$=-$\Omega(K')$. Consequently, the net Berry curvature integrated over the first Brillouin zone vanishes, precluding any valley-polarized transport. $\mathcal{T}$ symmetry must also be broken. In a $\mathcal{T}$ symmetric system, the Berry curvature obeys $\Omega(K)$=-$\Omega(-K)$. For hexagonal lattices, the $K$ and $K'$ valleys are linked by $\mathcal{T}$ symmetry, yielding Berry curvatures of equal magnitude but opposite sign, again resulting in a zero net Berry curvature. Thus, only when both $\mathcal{P}$ and $\mathcal{T}$ are broken can the symmetric constraint between the two valleys be lifted, allowing an asymmetric Berry curvature distribution and a finite net Berry curvature. Achieving a layer-locked AVHE further requires a band structure with clear layer resolution. Such a band structure enables carriers to acquire a layer-dependent anomalous velocity under a finite Berry curvature, inducing a lateral deflection within each layer. The bilayer $VS_2$ system simultaneously fulfills all these prerequisites. Interlayer sliding drives FE order, which breaks $\mathcal{P}$ symmetry. Interlayer AFM order, characterized by antiparallel magnetic moments in the top and bottom layers, breaks $\mathcal{T}$ symmetry. Moreover, the bilayer architecture inherently provides a layer-resolved band structure, satisfying the final requirement.

To systematically explore this platform, we consider four distinct magnetic configurations for the AA-1/AA-2 systems: P↑M↓↑, P↓M↓↑, P↑M↑↓, and P↓M↑↓, where P↑ (P↓) denotes FE polarization along +*z* (-*z*), and M↓↑ (M↑↓) represents the relative alignment of magnetic moments between the top and bottom layers [Fig. S7]. The spin-resolved band structures for the P↑M↓↑ and P↓M↓↑ states under SOC, which share the same AFM order but have opposite FE polarization, are compared in Figs. 3(a) and 3(d)]. These two states differ solely in the sign of the FE polarization, while sharing the same interlayer AFM order. Broken $\mathcal{P}$ symmetry and SOC lift the degeneracy of the *K* and *K′* valleys in the valence band, generating a spontaneous valley polarization, which we quantify as $\Delta V = E_K - E_{K'}$. In the P↑M↓↑ state, this yields layer-locked band-edge states: the VBM at *K′* is composed of top layer spin-down states, while the VBM at *K* originates from bottom layer spin-up states [Figs. 3(a) and 3(b)]. The resultant valley splitting of -68 meV provides a robust, electrically switchable platform for encoding valley-based information.

Reversible interlayer sliding switches the system between two stable FE states, an additional structural degree of freedom that couples directly to the valley and spin degrees of freedom. This coupling manifests as a FE valley-layer locking mechanism, whereby switching the FE polarization direction deterministically inverts the valley polarization. As shown in Figs. 3(d) and 3(e)], the reversal of FE polarization concurrently inverts the sign of valley polarization (68 meV), flips its associated spin channel, and swaps its layer of origin. This process effectively entangles the FE, valley, spin, and layer degrees of freedom.

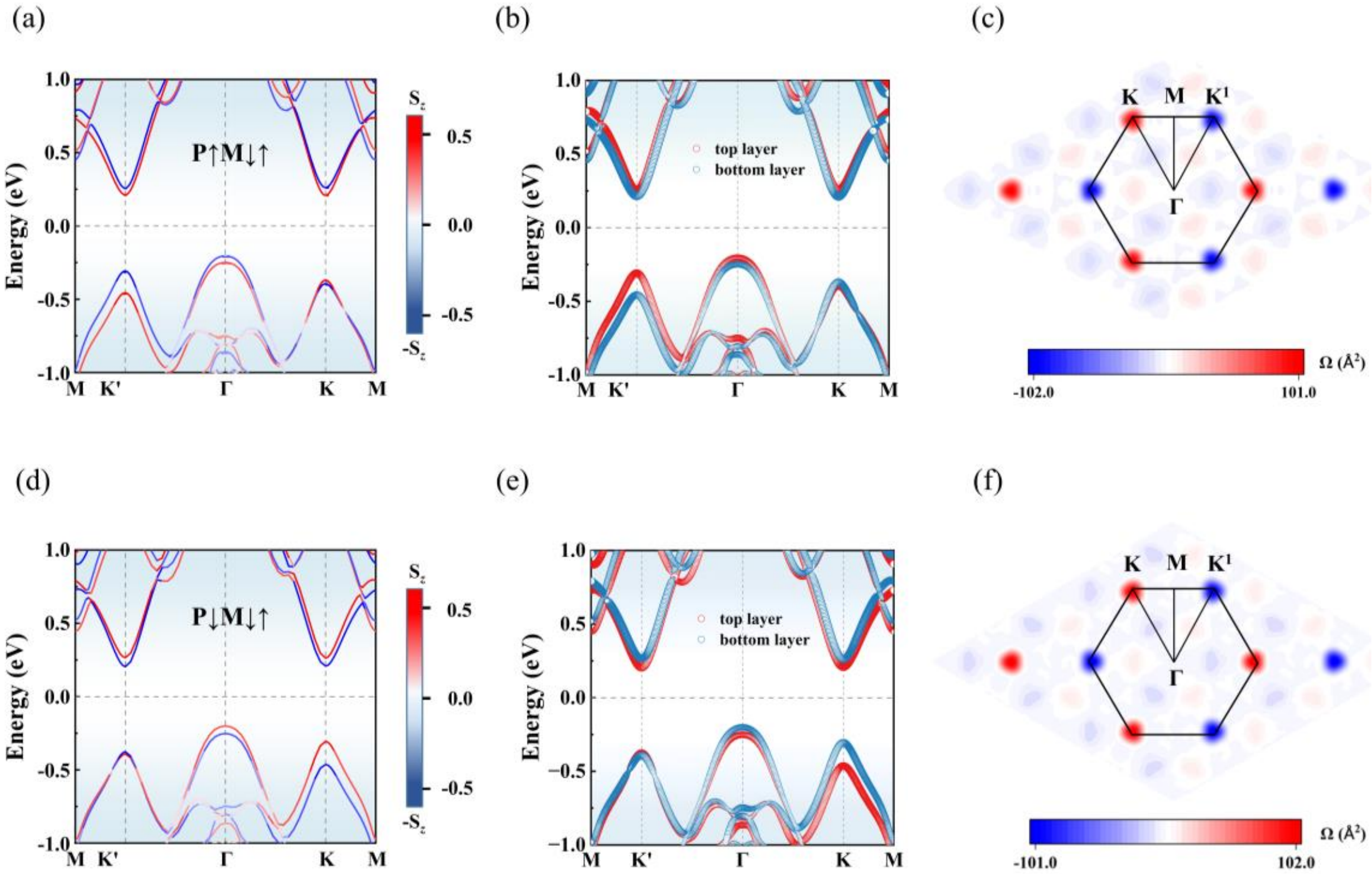

FIG. 3. Projected band structures for configurations (a) P↑M↓↑ and (d) P↓M↓↑, with colour indicating the spin projection along *z*-direction (red, spin-up; blue, spin-down). (b,e) Corresponding layer-projected band structures, showing contributions from the top (red) and bottom (blue) layers, respectively. (c,f) Momentum-resolved Berry curvature, for the (c) P↑M↓↑ and (f) P↓M↓↑ states.

To isolate the specific role of magnetic order, we analyze the P↑M↑↓ and P↓M↑↓ configurations. These share a common FE polarization relative to the first two states, but feature opposing interlayer magnetic alignments [Figs. 4(a) and 4(d)]. Crucially, while both exhibit valley polarization, their spin channels, valley splitting magnitudes, and the energy ordering of the valleys are all inverted with respect to their magnetic counterparts. For instance, in the P↑M↑↓ state, the VBM at *K* derives from top layer spin-up states, while at *K′* it arises from bottom layer spin-down states, yielding a valley polarization of +68 meV, identical in magnitude but opposite in sign to that of the P↑M↓↑ state. Thus, in this multiferroic bilayer, switching either the FE polarization or the magnetic order provides equivalent control over the valley-layer degree of freedom. These results establish a deterministic magnetoelectric mechanism for manipulating correlated electronic states and offer a design principle for valley and layer Hall devices in 2D magnets.

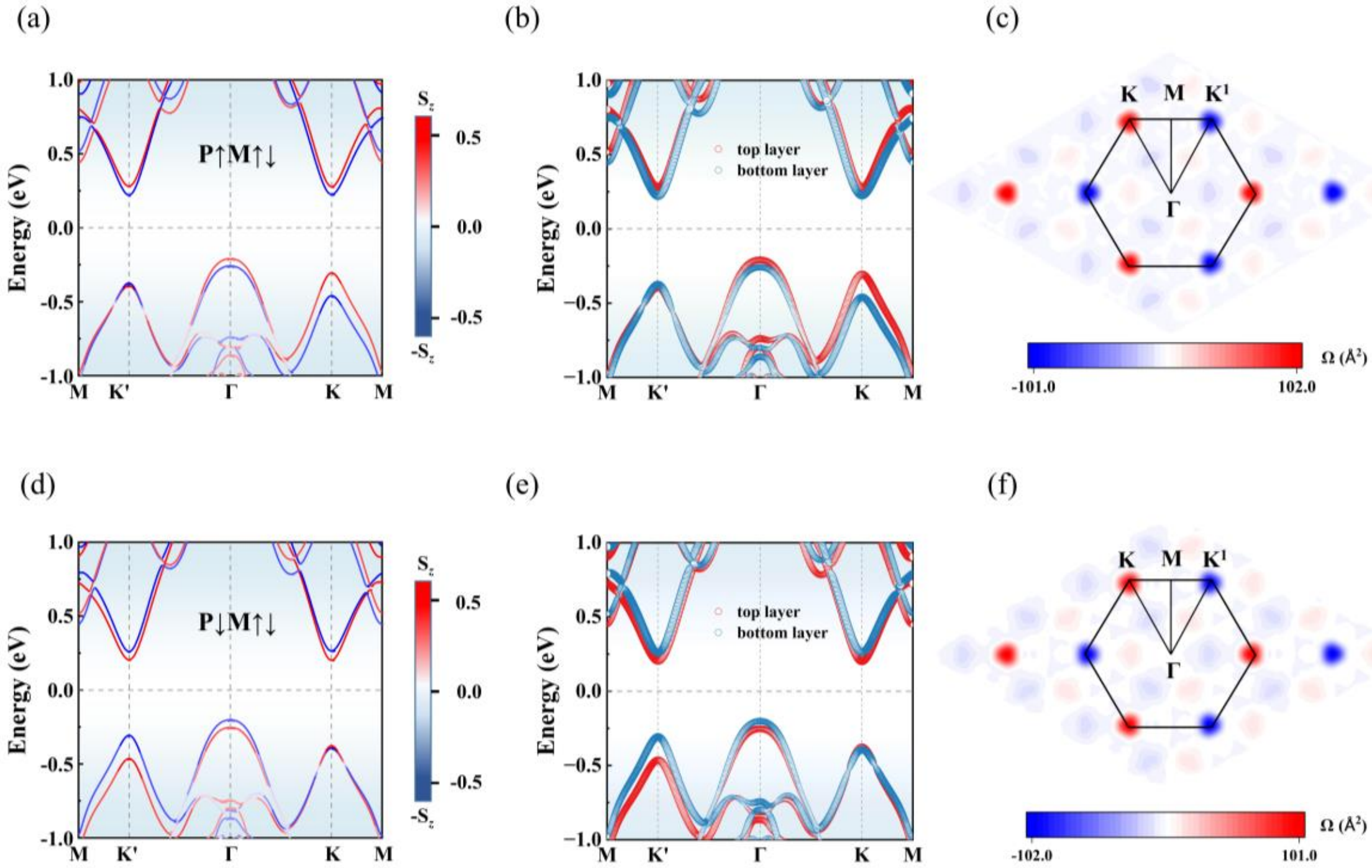

FIG. 4. Projected band structures for configurations (a) P↑M↑↓ and (d) P↓M↑↓. (b,e) Corresponding layer-projected band structures. (c,f) Momentum-resolved Berry curvature, for the (c) P↑M↑↓ and (f) P↓M↑↓ states.

The strong coupling between spin, valley, and layer degrees of freedom in this system locks the Berry curvature to the multiferroic order. To quantify this, we calculate the Berry curvature from the Kubo formula [59]:

$$\Omega_z(k) = -\sum_n \sum_{n\neq m} f_n(\vec{k}) \frac{2\,\mathrm{Im}\langle \psi_{nk} | \hat{\upsilon}_x | \psi_{mk} \rangle \langle \psi_{mk} | \hat{\upsilon}_y | \psi_{nk} \rangle}{(E_{nk} - E_{mk})^2} \qquad (1)$$

where $f_n(k)$ is the Fermi-Dirac distribution function, $\psi_{nk}$ ($\psi_{mk}$) the Bloch wave function with eigenvalue $E_{nk}$ ($E_{mk}$), $\hat{\upsilon}_{x/y}$ is the velocity operator along the *x*/*y* direction, respectively. In the P↑M↓↑ configuration, the *K* and *K′* valleys possess

Berry curvatures of 101 and -102 Å$^2$, respectively [Fig. 3(c)]. FE polarization reversal via interlayer sliding (P↑M↓↑→P↓M↓↑) switches these values to 102 and -101 Å$^2$, directly inverting the valley polarization [Fig. 3(f)]. Consequently, controlling the FE polarization direction through interlayer sliding provides deterministic electrical control over the valley polarization state. An equivalent inversion is achieved by flipping the magnetic order instead: the Berry curvatures in the P↑M↑↓ state (102 Å$^2$ at *K*, -101 Å$^2$ at *K'*) swap to 101 and -102 Å$^2$ in the P↓M↑↓ state [Figs. 4(c) and 4(f)]. This demonstrates a functional equivalence between FE polarization switching and magnetic moment reversal for controlling the valley-layer state. The result establishes deterministic, bidirectional control over the layer-locked AVHE via electric or magnetic fields, defining a multiferroic platform for valleytronics.

The Berry curvature functions as a momentum-space pseudomagnetic field. Its large magnitude at the valence band edges imparts a significant anomalous velocity to Bloch electrons in the presence of an in-plane electric field *E*, governed by the relation $v \approx -\frac{e}{h} E \times \Omega(k)$ [60]. In the P↑M↓↑ state, hole doping shifts the Fermi level into the valence band [Fig. 5(a)]. The layer-locked nature of the Berry curvature leads to a transverse deflection of spin-down *K'*-valley holes into the top layer, generating a layer-polarized Hall voltage, a direct signature of the layer-locked AVHE. Reversing the FE polarization to the P↓M↓↑ state inverts both the valley and layer polarization and the sign of the Berry curvature. Consequently, spin-up *K*-valley holes are deflected into the bottom layer, producing a layer-locked AVHE with opposite layer polarization [Fig. 5(b)]. Flipping the magnetic order achieves an equivalent result: in P↑M↑↓, spin-up *K*-valley holes accumulate in the top layer, while in P↓M↑↓, spin-down *K'*-valley holes accumulate in the bottom layer [Figs. 5(c) and 5(d)]. Notably, the layer-locked AVHE is observed in all four configurations defined by the combination of FE polarization reversal and magnetic moment reversal. This observation not only confirms the robustness of the layer-locked characteristics but also reveals two fundamental features: an intrinsic coupling among valley, layer and Berry curvature, and the functional equivalence of FE polarization reversal and magnetic moment reversal in modulating valley transport.

It should be noted that conventional non-magnetic FE materials can host a conventional valley Hall effect (VHE) through $\mathcal{P}$ symmetry breaking associated with the FE order. However, because they lack intrinsic magnetic order, they generally do not provide the internal $\mathcal{T}$ symmetry breaking needed for the coupled spin-valley-layer physics addressed here. In sharp contrast, the bilayer $VS_2$ system studied here exhibits simultaneous intrinsic breaking of both $\mathcal{P}$ and $\mathcal{T}$ symmetries. This combination naturally gives rise to a layer-locked AVHE with coupled valley, spin, and layer selectivity, as well as coordinated dual-switch control via electric and magnetic stimuli. Collectively, these characteristics establish a new paradigm for electromagnetically synergistic, multi-degree-of-freedom valleytronic control, laying a critical physical foundation for the design of next-generation valleytronic devices.

The high-symmetry AA-0 stacked bilayer provides a critical control case. Its centrosymmetric structure, protected by both mirror and $\mathcal{P}$ symmetry, precludes FE order. Accordingly, its band structure exhibits no valley polarization even under SOC [Fig. S8(a)]. While the layer-resolved Berry curvature can produce a layer-locked VHE, deflecting $K$ and $K'$ holes into opposite layers under an in-plane field [Fig. S8(b) and S8(c)], this response is symmetry-locked and non-switchable [Fig. S8(d)]. In contrast to the FE reversible, layer-locked AVHE in the AA-1/AA-2 systems, the absence of switchability in AA-0 confirms that FE polarization is essential for achieving electrically tunable valley-layer coupling.

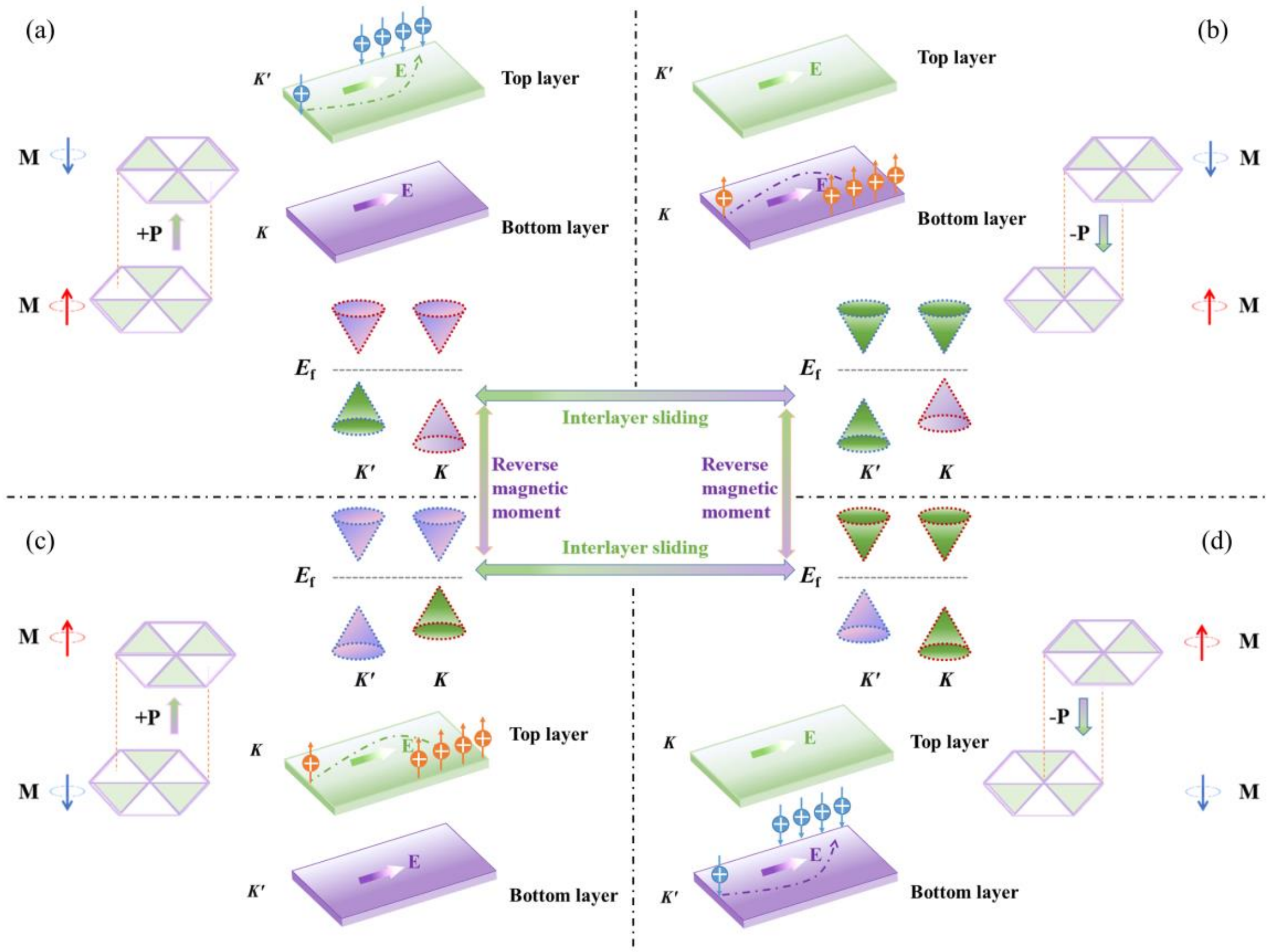

FIG. 5. Schematic illustration of the structure, valley-polarization, and layer-locked AVHE in hole-doped bilayer $VS_2$ for four distinct configurations: (a) P↑M↓↑, (b) P↓M↓↑, (c) P↑M↑↓, and (d) P↓M↑↓. In the valley-polarization diagrams, the spin-down and spin-up bands of the $K$ and $K'$ valleys are represented by blue and red dashed lines, respectively. The layer-resolved contributions are visualized by purple (lower layer) and green (upper layer) isosurfaces, with the Fermi level marked. In the AVHE schematics, holes (+) from the band edges are color-coded by their layer origin (purple: bottom; green: top), and circular arrows denote their spin direction, demonstrating the coupled spin-valley-layer selectivity of the transport.

## IV. CONCLUSION

In summary, our work establishes bilayer $VS_2$ as a 2D sliding FE-AFM, where interlayer sliding and magnetic reversal provide dual, equivalent pathways for controlling the valley-layer-spin landscape. First-principles calculations reveal that the symmetry breaking from sliding induces a switchable out-of-plane FE polarization, whose microscopic origin is an asymmetric interlayer charge transfer. This FE order couples directly to the AFM configuration, enabling deterministic control over valley polarization and a layer-locked Berry curvature. These interactions underpin the emergence of a fully electrically switchable and layer-locked AVHE. Crucially, we demonstrate a robust magnetoelectric coupling where FE polarization switching and magnetic moment flipping are functionally interchangeable mechanisms for toggling the layer-locked AVHE. This dual-switch mechanism, capable of electrically or magnetically addressing quantum states, defines a novel paradigm for controlling entangled electronic degrees of freedom in 2D multiferroics. More importantly, it establishes a concrete design principle for developing next-generation multi-state non-volatile memory and reconfigurable spin-valleytronic logic devices.

## SUPPORTING INFORMATION

See the supplementary material for the band structure and stability, magnetic configuration, and structural polytypes, etc.

## ACKNOWLEDGEMENTS

This work was supported by the National Natural Science Foundation of China (Grant Nos. 12364007, 12264016) and the Scientific Research Foundation of Hunan Provincial Education Department, China (Grant No. 24A0366). And the computing work is supported by the Open Source Supercomputing Center of S-A-I.